\documentclass{llncs}

\usepackage{graphicx}
\usepackage{amssymb}

\newcommand{\mygraphic}[1]{\includegraphics[height=#1]{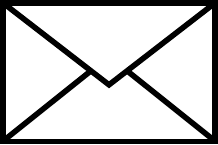}}
\newcommand{\myenv}{(\raisebox{0pt}{\mygraphic{.6em}})}
\newcommand{\myauthor}[1]{#1~\myenv}
\begin{document}

\title{A Security Framework for Cloud Data Storage(CDS) based on Agent}
\author{Oussama Arki\inst{1\myauthor{}} , Abdelhafid Zitouni\inst{1}}
\institute{$^{1}$ Lire Labs , Abdelhamid Mehri Constantine 2 University\\Ali Mendjli, 25000 Constantine , Algeria\\
 \email{$\lbrace$oussama.arki,abdelhafid.zitouni$\rbrace$@univ-constantine2.dz}}
\maketitle

\begin{abstract}
The Cloud has become a new Information Technology(IT) model for delivering resources such as computing and storage to customers on demand, it provides both high flexibility and resources use. However we are gaining these advantages at the cost of high security threats, which presents the major brake for the migration towards Cloud Computing.\\
Cloud Data Storage(CDS) is one of the Cloud services, it allows users to store their data in the Cloud, this service is very useful for companies and individuals, but data security remains the problem which makes customers worried about their data that reside in the Cloud. In this paper, we propose a framework of security to ensure the CDS, which is based on agents, it contains three layers: Cloud Provider layer, Customer layer and Trusted Third Party(TTP) layer.
\end{abstract}
\section{Introduction}\label{sec:Introduction}
In the last years, computing and storage technologies are rapidly developed, one of the major reasons is decreasing costs and increasing power of the computer resources beside to the success of Internet, which led to the situation where a big volume of data can be collected, stored and treated.\\
Cloud Computing is a new IT model, it  provides storage and computation resources as a service to the Cloud Customers (CC), often through a network (typically the Internet ).\\
Cloud Data Storage is one of the Cloud services. It allows users to store their data in the Cloud, by reserving a virtual space in the Cloud, it also provides the powerful way of managing data. Cloud Data Storage allows to store and manage the data remotely. Users do not have to buy the expensive hardware and to have policies to regulate and manage the data \cite{1}. The major brake for the adoption of this service is the data security. The virtualisation and the co-location of the physical resources are the principal characteristics that distinguish the Cloud Computing, which make the data security in the Cloud more difficult and complex than the traditional systems. The security of data in the Cloud is the biggest challenge of Cloud Providers (CP).\\
To ensure the CDS security and safety, we propose in this paper a framework of security, which contains three layers : the Customer layer, the Cloud Provider layer and the TTP layer. It is based on trust model and mobile agents.\\
This paper is organized as follows: Section two introduces Cloud Computing. Section three discusses the Information Security and Cloud Storage. Section four is about the related works. Section five presents our framework and the last section is a conclusion.
\section{Cloud Computing Overview}\label{sec:Cloud Computing Overview}
According to the famous definition of NIST (National Institute of Standards and Technology): Cloud Computing is a model for enabling ubiquitous, convenient, on-demand network access to a shared pool of configurable computing resources (e.g., networks, servers,storage, applications, and services) that can be rapidly provisioned and released with minimal management effort or service provider interaction \cite{2}.\\
This Cloud model is composed of five essential characteristics, three service Models, and four deployment models \cite{3} ,as in Figure 01.
\begin{figure}[h]
\begin{center}
\includegraphics[width=3.5in,height=1.9in]{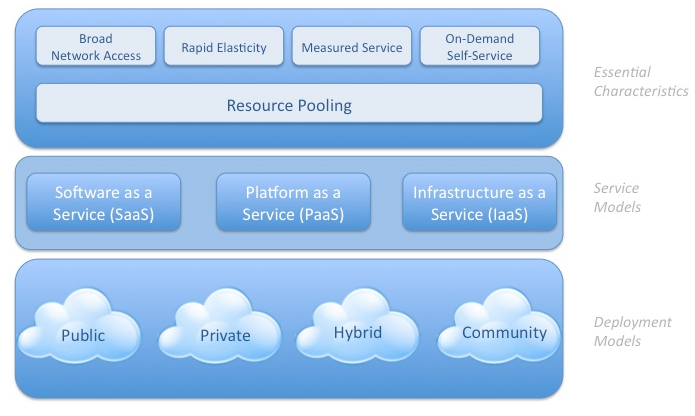} 
\caption{NIST Visual Model of Cloud Computing Definition \cite{4}}
\end{center}
\end{figure} 
\section{Information Security and Cloud Storage }\label{sec:Information Security and Cloud Storage}
In This section, we speak about Information Security in the Cloud and the Cloud Storage concerns.
\subsection{Information Security}
The term Information Security means protecting information and information systems from unauthorized access, use, disclosure, disruption, modification, or destruction in order to provide integrity, confidentiality and availability. Here, integrity means guarding against improper information modification or destruction, and includes ensuring information non-repudiation and authenticity. Confidentiality means preserving authorized restrictions on disclosure and access, including means for defending proprietary and personal privacy information. And availability means ensuring timely and reliable access to and use of information \cite{5}.\\
To understand this term in Cloud Computing environment, we must take into account the three main properties of Information Security \cite{6} :\\ 
\textbf {Confidentiality:} It means keeping user data secret in the Cloud systems. It ensures that user data which reside in Cloud cannot be accessed by unauthorized person. There are two basic approaches to achieve such confidentiality, physical isolation and cryptography. Confidentiality can be achieved through proper encryption technique: symmetric and asymmetric algorithms.\\
\textbf {Data Integrity:} It means Keeping data integrity is a fundamental task. It means in Cloud system is to preserve information integrity. Data could be encrypted to provide confidentiality but it will not guarantee that data reside on Cloud has not been altered. There are two approaches which provide integrity Digital Signature and Message Authentication Code.\\
\textbf {Availability:} Data should be available when it is requested via authorized user. It ensure that user can be able to use the service any time from any place.Two strategies called Hardening and Redundancy are mainly used to enhance the availability.
\subsection{Cloud Storage}
Cloud storage is an important service of Cloud Computing, which offers services for data owners to host their data in the Cloud. This new paradigm of data hosting and data access services introduces two major security concerns \cite{3}:
\begin{itemize}
\item \textbf{Protection of data integrity}. Data owners may not fully trust the Cloud server and worry that data stored in the Cloud could be corrupted or even removed.
\item \textbf{Data access control}. Data owners may worry that some dishonest servers give data access to unauthorized users, such that they can no longer rely on the servers to conduct data access control .
\end{itemize}
\section{Related works}\label{sec:Related works}
In the last years, many researchers used agents in their works, to ensure the security in Cloud Computing.\\
In \cite{7} Shantanu \& al, proposed a Cloud security two-tier framework, which is based on a trust model and the use of agents, they used two agents for the trust, one in Customer side and the other in Cloud Provider side, these agents analyse the user behavior, to ensure that the user remains always as a trusted entity before the interaction with the Cloud Provider. This model suffers from some weakness, because there is no mechanism to prevent malicious activity without Cloud service provider information about user activities, and the use of proxy server, which presents also a weak point, if it crashes, the customer can not communicate with the Cloud Provider.\\
In \cite{8} Priyank \& al, proposed an MAS framework to ensure the security of the customer resources in the Cloud, they used mobile agents to supervise the hyper-visor and to collect information from the virtual machines of the customer, these agents monitor the virtual machines, to keep their privacy and integrity .If there is any problem, they notify it. The weak point in this model is that the user side is ignored, like the user identification and the identity management.\\
In \cite{9} Amir \& al ,proposed a framework of MAS to facilate security of Cloud Data Storage, the proposed framework has two layers: Customer layer and Cloud Data Storage layer. In the Customer layer an interface agent is used for the interaction with the customer, the other agents are distributed in the Cloud Data Storage layer, each one of them has a specific task and they communicate to achieve the global goal. The major problem in this work is that the security of the Customer layer is not dealt with, specialy the mechanism for the identification of the customer.\\
In \cite{10} Alwesabi \& al, proposed a MAS framework of Cloud security in the form of two-tier framework : the Virtual Server layer and the Cloud Provider layer. The communication between the two layers is ensured by the use of mobile agents. In the Virtual Server side they used an analyzer agent for the  authentication of the customers, and in the Cloud Provider side a security agent is used to ensure the security, the problem in this framework is that the security mechanism used by the security agent is not clear.\\
In \cite{11} Benabied \& al, proposed a MAS framework to make safe Cloud environment, which contains two layers : Cloud Provider layer and the Customer layer. The communication between the two layers is ensured by the use of mobile agents, to ensure the trust at Cloud environment, this framework based on the use of Trust Model, they used two types of agents, one in Customer side and the other in Cloud Provider side, this model ensures the customer trust and  guarantees that only the trusted customers  can interact with the Cloud Provider. The weakness of this framework is  the monitoring of the Cloud Provider side ,which is ignored, like the supervising  of the virtual machines.\\
In \cite{12} Md. Rafiqul \& al, proposed an agent based framework for providing security to Data Storage in Cloud, which contains three levels: Customer level, Cloud Provider level and Data Storage level. In their work, they considered that the degree of data security is depended on data sensitivity, because of that, they classified the sensitivity of data into five categories, for each category they used a different method of authentication, a different algorithm of encoding and a different function of hashing. By using this technique they increased the Cloud performance, but they risk the data of the customer.\\
In \cite{13} Venkateshwaran \& al, proposed a security framework for Agent-Based Cloud Computing, they based on the use of Trust Model. In their work, they used a security agent, which ensures the security, it is responsible for the authentication of the customers and the analysis of the customer's trust. By using this Trust Model, they guarantee that only trusted users can interact whith the Cloud Provider .

\section{The Proposed Framework}\label{sec:The Proposed Framework}
To ensure the security of data in the Cloud, we proposed a framework of security based on agents, which contains three layers: the Customer layer, the Cloud Provider layer and the TTP layer. In this work, we used a Trust Model to manage the confidence of the customers, an Encryption Method to ensure the privacy and the confidentiality of data and an Integrity Technique to allow the user to check the integrity of his data in the Cloud.\\
The TTP Layer is an intermediary between the Customer layer and the Cloud Provider layer. It is responsible for querying the Cloud Provider on behalf of user. It has expertise and capabilities that users may not have \cite{1}. It is responsible for the Encoding/Decoding operation of data, before storing them in the Cloud. It is also used to check the integrity of data, without recovering them by the customer.
Figure 02 presents our framework .
\begin{figure*}[h]
\begin{center}
\includegraphics[height=4in,width=4.8in]{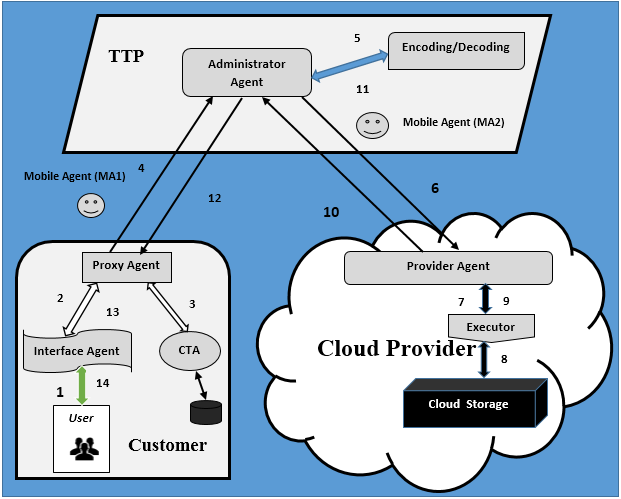} 
\caption{Framework Description}
\end{center}
\end{figure*}
\subsection{Framework Description }
To store his data in the Cloud, the customer must initially authenticate himself, for that it provides his information of identity to the Interface Agent. This last  analyzes  the information seized by the customer, then asks the Proxy Agent to check the identity of the customer. The Proxy Agent checks the identity of the customer through his database. If the customer is well authenticated, the Proxy Agent requests the Interface Agent to post the user interface to the customer, and asks the Customer Trust Agent(CTA) to calculate the trust degree of this customer. Only if the trust degree of the customer is enough to establish a connection with the  Cloud, the Proxy Agent creates a Mobile Agent(MA1), to carry the request of the customer towards the TTP layer.\\
When the first phase is well established, if the customer wants to transfer his data towards the Cloud ,the Administrator Agent obtains the data from MA1, which came from the Customer layer, and gives them to the Encoding Agent, which encodes them, to ensure that no one can read the data that will be transferred towards the Cloud (even for the Cloud Provider). After the encoding operation, the Administrator Agent creates another Mobile Agent(MA2), to carry the data towards the Cloud Provider .\\
When the MA2 arrives at the Cloud layer, it contacts the Provider Agent. If all is regulated on the Storage level, the Provider Agent asks the Executor Agent to execute the request and to store the data of the customer.\\
In the opposite direction, to turn over the data to the customer, the same procedure is carried out , the Provider Agent gives the data that is retrieved by the Executor Agent to the MA2, that came from the TTP layer, which moves towards the TTP layer, then the Encoding Agent decodes the data of the customer, after the decoding operation, the Administrator Agent gives the data to the MA1 that came from the customer layer, which moves to the Customer layer and gives the data to the Proxy Agent, this last gives them to the Interface Agent to display them to the customer.\\
The communication between the Customer layer and Cloud Provider layer through the TTP layer, can be summarized with the two sequence diagrams below, where the steps are as follows:
\begin{itemize}
\item \textbf {Step 01:} The customer sends his information of authentication to the Interface Agent, which analyzes them, if the information is normal, it sends them to the Proxy Agent.
\item \textbf {Step 02:} The Proxy Agent checks the identity of the Customer, if the information is correct, it connects himself to the Interface Agent with SSL connection and asks him to post the user interface to the customer. If it is not, the access is rejected.
\item \textbf {Step 03:} The Proxy Agent requests the consumer trust degree from the CTA. Which calculates the trust degree of the customer and indicates if the customer is a trusted entity or not.
\item \textbf {Step 04:} The Proxy Agent receives the response from the CTA, if the customer's trust degree is more than the threshold for the connection, it creates the MA1 , which carries the request of the customer towards the TTP layer. If it is not, the request is removed .  
\item \textbf {Step 05:} The Administrator Agent in the TTP layer receives the MA1, if the customer wants to store his data, it passes the data of the customer to the Encoding Agent, which encodes them and turns over them towards the Administrator Agent.
\begin{figure}[h]
\begin{center}
\includegraphics[height=2.6in,width=4.2in]{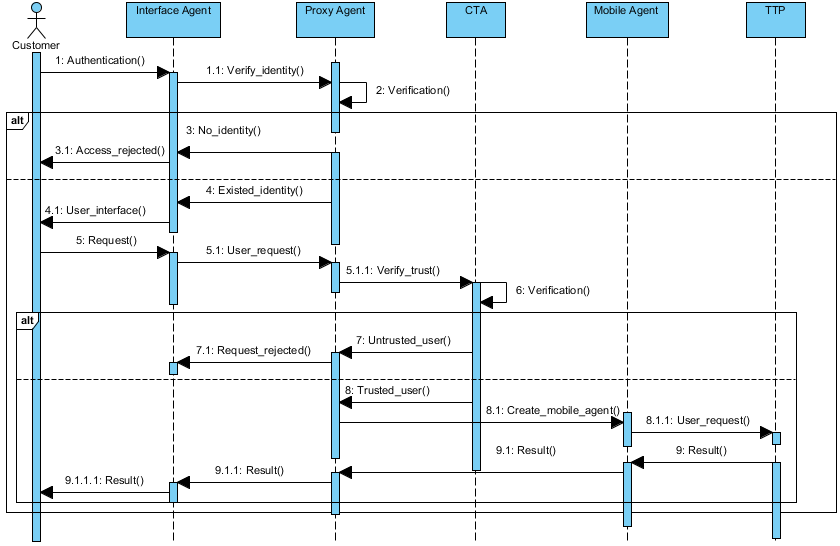}
\end{center}
\caption{Interaction Customer/TTP}
\end{figure}
\item \textbf {Step 06:} The Administrator Agent creates MA2, which carries the request of the customer towards the Cloud Provider layer.
\item \textbf {Step 07:} The Provider Agent receives the MA2, and asks the Executor Agent to carry out the request of the customer.
\item \textbf {Step 08:} The Executor Agent carries out the request of the customer (for storing the data or to turn over them).
\item \textbf {Step 09:} If the customer asked to recover his data, the Executor Agent turns over them and gives them to the Provider Agent.
\item \textbf {Step 10:} The Provider Agent passes the data to the MA2, which moves towards the TTP layer and gives them to the Administrator Agent.
\item \textbf {Step 11:} The Administrator Agent asks the Encoding Agent to decode the data of the customer. This last decodes them, and turns over them to the Administrator Agent.
\item \textbf {Step 12:} The Administrator Agent gives the data to the MA1, which moves towards the Customer layer and passes them to the Proxy Agent.
\item \textbf {Step 13:} The Proxy Agent receives the data and gives them to the Interface Agent.
\item \textbf {Step 14:} The Interface Agent displays the data to the customer, and the customer can recover his data in full security.
\end{itemize}

\begin{figure}[h]
\begin{center}
\includegraphics[height=2.6in,width=4.2in]{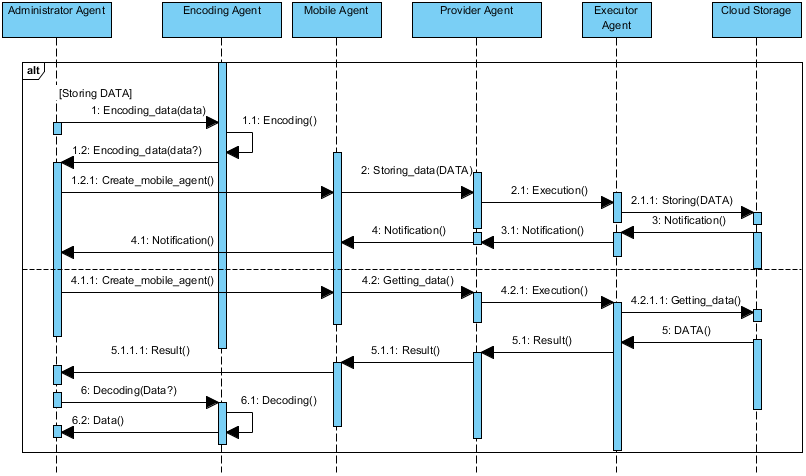}
\end{center}
\caption{Interaction TTP/Cloud Provider}
\end{figure}
\subsection{Trust Evaluation}  
Trust is one of the most important means to improve security and privacy of Cloud platforms. While in fact trust is the most complex relationship between entities because it is extremely abstract, unstable and difficult to be measured and managed \cite{14}. In our framework, we based on the work of \cite{15}, which is a pervasive trust management model for dynamic open environments. It based on fuzzy logic to calculate the trust degree of the customer, according to the actions carried out by them.\\ 
To calculate the trust degree of the customer, we based on the continuous examination of the customer behavior, the CTA uses a function of trust to indicate the trust degree of each customer. After each action doing by the customer, this agent recalculates the trust degree of the customer. The customer trust degree can be reduced or increased by the CTA according to actions carried out by the customer. At the end of each evaluation the customers are classified in three categories: trusted entity, innocent entity and untrusted entity  \cite{15}.\\
We can also classify the actions carried out by the customers in two classes : positive actions and negative actions, the positive actions are correct actions carried out by the trusted entity, However, we assume that all negative actions are not the same, that is the reason because we distinguish between wrong actions and malicious actions, the wrong actions are bad actions which do not cause any damage to the system as the attempt to access to unauthorized resources, they are carried out by the innocent entity, but the malicious actions are harmful actions like attacks, they are carried out by the untrusted entity \cite{15}.\\
To calculate the action value Pa, we take into account the performed action weight, but this value is penalised or rewarded by the past behaviour. This function increases or decreases according to the performed positive and negative actions respectively, so \cite{15}:
\[Pa =\left(1-\frac{Na}{Totala} \right) .Wa^{m} \      \quad \quad where \quad 0\textless= Pa\textless=1  \]    
- \textbf {(1-(Na/Totala))}: means the past behaviour of the customer, it goes towards  \textbf {0} if the behavior is negative, and towards \textbf {1} for a positive behavior .\\
-\textbf { Na}: is the number of negative actions realized by the customer.\\                               
-\textbf { Totala}: is the total number of performed actions by the customer during the interaction with the Cloud Provider.  \\     
-\textbf { Wa}: is the action's weight according to it's nature (positive, wrong, and malicious). Wa for positive actions is \textbf {1}, for wrong actions is \textbf {0.5} and for malicious actions is \textbf {0} .\\     
-\textbf { m}: the parameter m is the security level, where\textbf { m\textgreater=1}.\\			
When the customer realizes a new action, Pa is recomputed, which reflects the present behavior of the customer, the new trust degree will take it into account and it will modify the current trust degree.\\
By the use of this model in our framework, we have an efficient method for the access management, we guarantee that only the authorised customers can interact with the Cloud Provider, we have also a kind of monitoring of the actions carried out by the customers. So we can stop any attempt of malicious action like the access for unauthorized resources, abuse of Cloud services or any attacks.
\subsection{Data Encryption}
For the data encryption phase, there is two steps, to guarantee the data confidentiality and privacy. The step of fragmentation, where we divide the data into many fragments, then a step of encoding, where RSA algorithm is used to find out the keys, these keys are used to encode and decode the file \cite{16}.\\
So the Encoding Agent divides initially the data of the customer into many fragments, then it encodes them by using the RSA algorithm. After the Encoding Operation, the TTP sends all of the fragments to the Cloud Provider to store them. So the Cloud Provider contains only a parts of data, which have no means. With this technique we guarantee that only the TTP can recover the data of the customer. Figure 05 show the Encoding Operation.
\begin{figure}[h]
\begin{center}
\includegraphics[height=1.4in,width=3.5in]{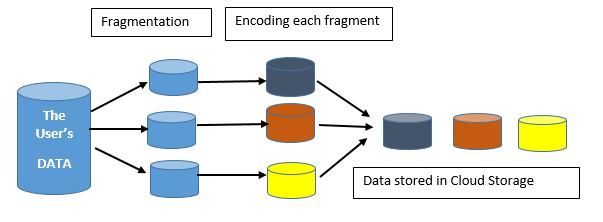}
\end{center}
\caption{Encoding Operation}
\end{figure}\\
For the retrieving of the data, the TTP carries out the opposite operation, to recover the real data of the customer from the fragments retrieved from the Cloud Provider. As in Figure 06.
\begin{figure}[h]
\begin{center}
\includegraphics[height=1.4in,width=3.5in]{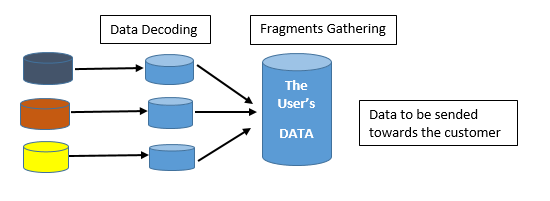}
\end{center}
\caption{Decoding Operation}
\end{figure}\\
With the encryption of the file at the TTP layer, we guarantee that the data of the customers are safe, so the client will not worry about the privacy and the confidentiality of his data, which reside in the Cloud.
\subsection{Data Intergrity check}
Integrity, in terms of data security, is nothing but the guarantee that data can only be accessed or modified by those authorized to do so, in simple word it is a process of verifying data. Data Integrity is very important among the other Cloud challenges. As data integrity gives the guarantee that data is of high quality, correct, unmodified \cite{17}.\\
To allow the customer to check the integrity of his data stored in the Cloud, we based on the Provable Data Possession(PDP) Scheme based on MAC(Message Authentication Code), to ensure data integrity of file F stored on Cloud storage in very simple way. The TTP computes a MAC for each fragment of the whole file with a set of secret keys and stores them locally before outsourcing it to the Cloud Service Provider(CSP). It Keeps only the computed MACs on his local storage, then sends the fragments of the file to the CSP, and deletes the local copy of the fragments. Whenever a customer needs to check the Data integrity of file F,the TTP sends a request to retrieve the file from CSP, reveals a secret keys to the CSP and asks to recompute the MACs of the fragments of the file, and compares the re-computed MACs with the previously stored values \cite{17}. \\
\begin{figure}[h]
\begin{center}
\includegraphics[height=1.3in,width=3in]{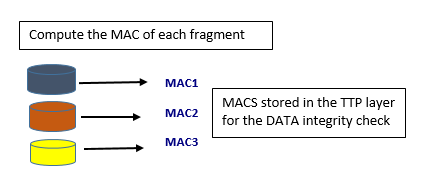}
\end{center}
\caption{MACs Computing}
\end{figure}\\
If the re-computed MACs at Cloud Provider and the MACs stored in the TTP are the same, then the data integrity is checked.
\section{Conclusion}\label{sec:Conclusion}
Cloud Computing security is very important for the continuity of this model, particularly in Cloud Storage service. In this paper we presented our proposed framework to ensure the security of Cloud Data Storage(CDS), which contains three layers. It is based on the use of agents, a Trust Model and the TTP which is used for the encryption and the integrity check of data.\\
Work is currently going on the framework implantation, where it will be applied to a specific case study. Further research could be realized to improve and to extend the present work, by including cognitive agents to make the interaction between these three layers more automatic, to increase Cloud performance.

%\section*{Acknowledgments}\label{sec:Acknowledgments}

%\bibliographystyle{unsrt} 
%\bibliography{biblio}

\end{document}